\documentclass[aps,prl,preprint,superscriptaddress]{revtex4-1}

\usepackage{graphicx, subfigure, amsmath}

\begin{document}

%\preprint{}

\title{Internal polarization dynamics of vector dissipative-soliton-resonance pulses in normal dispersion fiber lasers}
\affiliation{Jiangsu Key Laboratory of Advanced Laser Materials and Devices, Jiangsu Collaborative Innovation Center of Advanced Laser Technology and Emerging Industry, School of Physics and Electronic Engineering, Jiangsu Normal University, Xuzhou, Jiangsu 221116 China}
\affiliation{Department of optical science and engineering, Fudan University, Shanghai 200433, China}
\affiliation{Optical Sciences Group, Research School of Physics and Engineering, The Australian National University, Canberra ACT 2600, Australia}

\author{Daojing Li}
\affiliation{Department of optical science and engineering, Fudan University, Shanghai 200433, China}
\affiliation{Optical Sciences Group, Research School of Physics and Engineering, The Australian National University, Canberra ACT 2600, Australia}

\author{Deyuan Shen}
\affiliation{Department of optical science and engineering, Fudan University, Shanghai 200433, China}

\author{Lei Li}
\affiliation{Jiangsu Key Laboratory of Advanced Laser Materials and Devices, Jiangsu Collaborative Innovation Center of Advanced Laser Technology and Emerging Industry, School of Physics and Electronic Engineering, Jiangsu Normal University, Xuzhou, Jiangsu 221116 China}

\author{Dingyuan Tang}
\affiliation{Jiangsu Key Laboratory of Advanced Laser Materials and Devices, Jiangsu Collaborative Innovation Center of Advanced Laser Technology and Emerging Industry, School of Physics and Electronic Engineering, Jiangsu Normal University, Xuzhou, Jiangsu 221116 China}

\author{Lei Su}
\affiliation{School of Engineering and Materials Science, Queen Mary University of London, London, UK}

\author{Luming Zhao}
\email[]{zhaoluming@jsnu.edu.cn}
\affiliation{Jiangsu Key Laboratory of Advanced Laser Materials and Devices, Jiangsu Collaborative Innovation Center of Advanced Laser Technology and Emerging Industry, School of Physics and Electronic Engineering, Jiangsu Normal University, Xuzhou, Jiangsu 221116 China}
\affiliation{Key Laboratory of Optoelectronic Devices and Systems of Ministry of Education and Guangdong Province, Shenzhen University, 518060, China}

\date{\today}

\begin{abstract}
 Investigation of internal polarization dynamics of vector dissipative-soliton-resonance (DSR) pulses in a mode-locked fiber laser is presented. Stable vector DSR pulses are experimentally observed. Using a waveplate-analyzer configuration, we find that polarization is not uniform across a resonant dissipative soliton. Specifically, although the central plane wave of the resonant dissipative soliton acquires nearly a fixed polarization, the fronts feature polarization states that are different and spatially varying. This distinct polarizaiton distribution is maintained while the whole soliton structrue extends with varying gain conditions. Numerical simulation further confirms the experimental observations.
 \end{abstract}

% \pacs{}
%\keywords{}

\maketitle

Dissipative solitons are self-localized structures that appear widely in physical, biological, and medical systems with the presence of continuous energy or matter supply and dissipation \cite{ds:2005, Akhmediev:2013iw, ds:fromoptics,Herr:2013ip,Grelu:2012gc}. A very important aspect of the concept of the dissipative soliton is that it extends conventional optical soliton theory to nonlinear integrable systems and promotes the generation of ultrafast optical pulses in mode-locked fiber lasers with unprecedented energy levels \cite{Grelu:2012gc}. Particularly, a recently developing approach known as the dissipative-soliton-resonance (DSR) effect \cite{Akhmediev:2008hg,Chang:2008in,Chang:2009ga,Ding:2011ez}, provides the possibility of avoiding pulse break-up and has boosted   pulse energy impressively to tens of microjoules \cite{Du:2016ol,Semaan:2016ol,Zhao:2016ij,ArmasRivera:2016gp}.

Apart from its capacity for fairly large energy pulse generation, resonant dissipative solitons (also refered to as DSR pulses, interchangeably), also demonstrate distinctive behaviors of strong physical interest. A DSR pulse is a complex of two interacting fronts \cite{Anonymous:1990vy,Akhmediev:2008hg,Chang:2008in,Chang:2009ga}. At small energy supply, two fronts are closely connected, forming a regular dissipative soliton. With energy supply increasing, the resonance effect allows for two fronts moving apart from each other while it generates a plane wave in the center, which strongly binds two fronts together. The central plane wave extends, and the distance between the two fronts grows linearly and indefinitely with energy supply, while the fronts themselves do not change. The central plane wave and the fronts were found to feature different chirps: a moderately low linear chirp throughout the extended central plane wave and large linear chirps across both fronts \cite{Ding:2011ez,CuadradoLaborde:2016gm,Li:2015kxa,Li:2016sr}.

Internal soliton dynamics continually intrigues researchers due to its physical and practical interests. Here we carry out further research on internal polarization dynamics of the DSR pulse, taking into account its vector nature. There is a rich variety of vector soliton dynamics that has been predicted and observed in optical fibers and mode-locked fiber lasers \cite{AKHMEDIEV1994278,PhysRevE.51.3547,polmultiplex1992,Barad:1997cp,Islam:90,PhysRevLett.82.3988,Tang:2008cv, Zhao:2009fs,Zhang:08,Zhang:2009jd,Tang:2010hf,Sergeyev:2014dh,Sergeyev:12,Luo:2013fn}. A conventional soliton that forms as a balance of fiber dispersion and nonlinearity was found to acquire a single uniform state of polarization (SOP) \cite{AKHMEDIEV1994278,PhysRevE.51.3547,polmultiplex1992,Islam:90,Barad:1997cp}. During propagation in fibers, the SOP of the soliton either rotates as a unit or is locked. Dissipative solitons differ significantly from conventional solitons, since they require not only a balance of dispersion and nonlinearity, but also a balance of energy exchange \cite{Grelu:2012gc}. A considerable number of studies on vector dissipative solitons have focused on the stability and evolution of the solitons from roundtrip to roundtrip \cite{Sergeyev:2014dh,Sergeyev:12,Zhang:2009jd,Luo:2013fn,Tang:2010hf}. However, few insights into the internal polarization dynamics of dissipative solitons have been given. The duration of a dissipative soliton in a fiber laser ranges typically from femtoseconds to a few tens of picoseconds. Given the speed limitation of optical polarimeters, it is very challenging to gain insight into the internal polarization dynamics of  such a fast physical phenomenon. In this paper, we propose a qualitative investigation of the internal polarization dynamics of dissipative solitons in a mode-locked fiber laser where the DSR was reached. The resonance effect allows wider soliton generation, which facilitates the investigation. Using a waveplate-analyzer configuration and high-speed detection equipment, we are able to gain insight into the internal polarization dynamics of a dissipative soliton for the first time. Experimental results reveal that the central plane wave of the resonant dissipative soliton acquires nearly a fixed SOP. However, the fronts feature different and spatially varying SOP. This distinct polarization distribution is maintained while the whole soliton structure expands with increasing gain. Numerical simulation confirms the experimental observations.

\begin{figure}[tbp]
\centering
\includegraphics[width=2.7in]{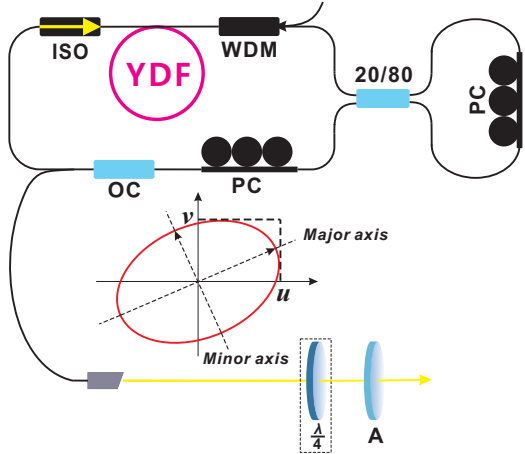}
\caption{Schematic of the laser setup. WDM: wavelength division multiplexer; ISO: isolator; PC: polarization controller; YDF: ytterbium-doped fiber; SMF: single-mode fiber; OC: output coupler; $\frac {\lambda} {4}$: quarter-wave plate; A: analyzer.}
\label{fig:setup}
\end{figure}

\begin{figure}[htbp]
\centering
\includegraphics[width=3.4in]{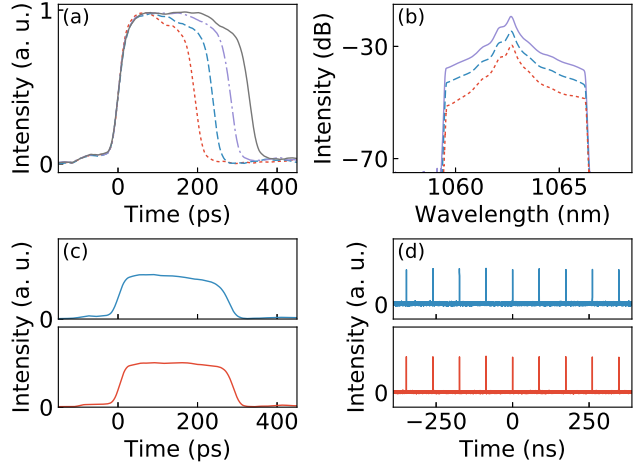}
\caption{(a) Single pulse traces as a function of the pump power, from innermost to outermost lines, 250, 300, 350, and $400~mW$. (b-d) Polarization-resolved measurements of the pulse under pump power of 350 $mW$ . (b) Spectra of original pulse (purple solid), major component (blue dashed), minor component (red dotted). (c) single pulse traces and (d) pulse trains of the major component (top row) and minor component (bottom row).}
\label{fig:2}
\end{figure}

The laser is shown in Fig.~\ref{fig:setup}. The gain was provided by a 42 $cm$ ytterbium-doped fiber (YDF), counter-pumped by a 976 $nm$ laser diode. A polarization-independent isolator ensures unidirectional operation. A 80:20 coupler and a 7-meter-long SMF were connected, thus forming a NOLM to provide saturable absorption. The split ratio of the coupler and the length of loop were chosen so as to ensure strong peak-power-clamping effect. This is crucial for achieving DSR. Furthermore, no narrow-band spectral filter was inserted to facilitate the DSR formation \cite{Li:2015kxa}. Two polarization controllers (PCs) were placed in the cavity, providing careful optimization of the cavity birefringence. The laser was outputted from a 30\% coupler and then monitored by a high-speed photodetector-oscilloscope combination with a bandwidth of 45 $GHz$.

By appropriate adjustment of the PCs, resonant dissipative solitons could be easily obtained in the laser. The pulse evolves into a flat top profile with increasing pump [Fig.~\ref{fig:2}(a)]. Pulse duration and energy show linear growth with pump power; these are typical DSR features. For polarization-resolved measurements, the laser output was connected to a collimator and then passed through a rotatable analyzer. The axis of the analyzer was aligned to the maximum (minimum) of the energy transmission, thus obtaining the major (minor) component of the pulse. We set a proper pump power of 350 $mW$ and conducted the polarization-resolved measurement. The original pulse is of 300 $ps$ width [purple dot-dashed line in Fig.~\ref{fig:2}(a)]. The ratio between maximum and minimum transmitted energy measured is around 12:1, indicating that the resonant pulse is elliptically polarized, i.e. it is a vector soliton.

Results of polarization-resolved measurement are summarized in the rest of Fig.~\ref{fig:2}. The spectra have steep edges and the same behavior with gain variation as was found in \cite{Li:2016sr}. Thus, with increasing pump, the newly-emerging spectrum appears at the center and forms a spike, while its rectangular pedestal remains unchanged. No wavelength shifting between two components is observed, suggesting a small cavity birefringence. Two components shows no evolution in pattens through roundtrips [Fig.~\ref{fig:2}(d)]. The radio frequency spectra of the components show no polarization evolution frequency \cite{PhysRevLett.82.3988}. These observations indicate that the laser generates stable vector DSR pulses.

To obtain more insight into the internal polarization properties of the DSR pulses, we inserted a quarter-wave plate before the rotating analyzer. We recall that for an elliptically-polarized light wave, the resolved components along the major and minor axes have a phase difference of $\pm \pi/2$. When the light wave passes through a quarter-wave plate, if the fast (or slow) axis of the quarter-wave plate is parallel to one of the principal axes of the incident light, the quarter-wave plate will induce a phase shift of $\pm \pi/2$ between the two components. The resulting phase shift between the major and minor components then becomes 0 or $\pm \pi$. Thus, the elliptically-polarized light will be converted to linearly-polarized light, which then can be blocked by a rotating polarizer at a certain angle.

\begin{figure}[tbp]
\centering
\subfigure {
    \includegraphics[width=2.in]{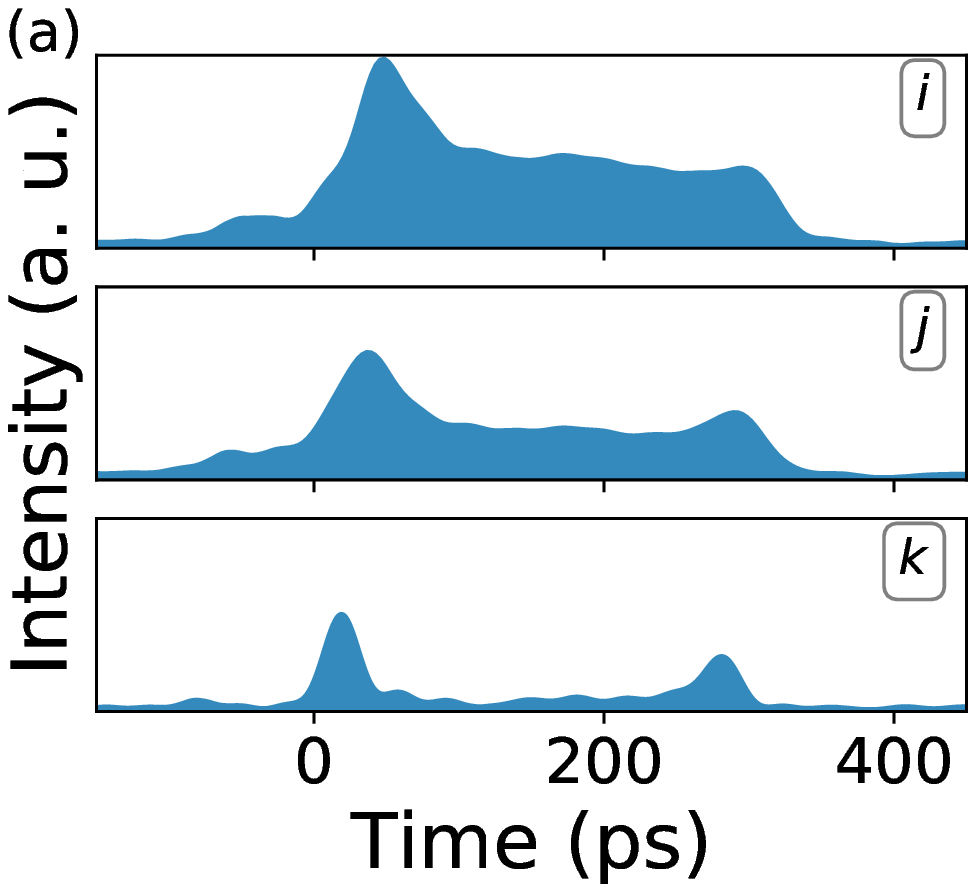}
    \label{fig:3a}
}
\hspace{0in}
\subfigure {
    \includegraphics[width=2.in]{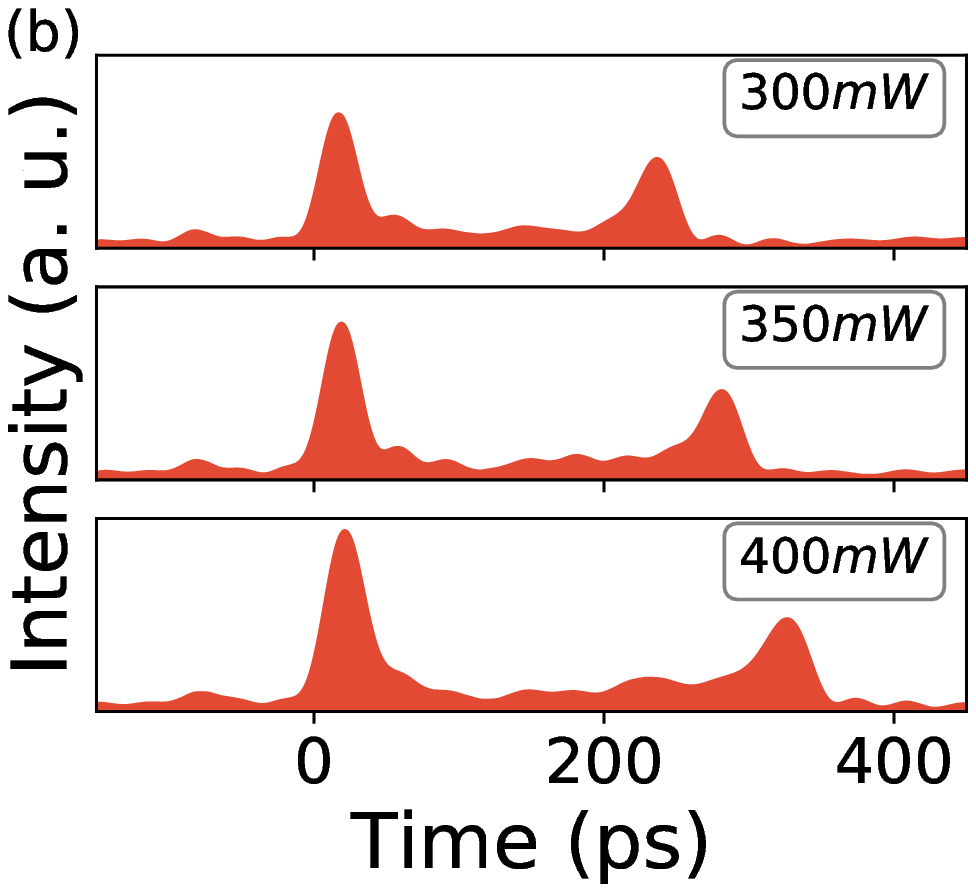}
    \label{fig:3b}
}
\caption{(a) Oscilloscope traces of pulse after the analyzer, \emph{i, j, k} represent different states with continuously-rotating analyzer. (b) Oscilloscope traces of pulse after the analyzer, versus increasing pump power, with analyzer fixed at state $k$. }
\label{fig:3}
\end{figure}

Accordingly, we used the waveplate-analyzer combination to test polarization dynamics of the DSR pulse. The axis of the analyzer was first set to match the minor axis of the DSR pulse by identifying the minimal energy transmission. Then a quarter-wave plate was inserted before the analyzer, with its fast axis carefully aligned parallel to the axis of the analyzer, therefore, parallel to the minor axis of the DSR pulse. With this set-up, if the incident pulse acquires a uniform elliptical polarization state, the pulse would be converted into a linearly-polarized one by the quarter-wave plate. Then, when rotating the following analyzer, linearly-polarized light could be fully blocked by the analyzer as a unit at a certain angle. Experimentally, we kept monitoring the transmitted light field while carefully rotating the analyzer. Traces \emph{i-k} in Fig. \ref{fig:3a} show different states with a continuously rotating analyzer axis in the vicinity of the minimum transmitted light field. It turns out that the central plane wave of the DSR pulse rises or falls together with the rotating analyzer, whereas the two fronts present asymmetric oscillatory structures. Eventually, the central part would decrease to almost zero. However, the fronts would not do this, as one can see from trace \emph{k} in Fig.~\ref{fig:3a}. It looks as if the fronts form two separated pulses. Rotating the analyzer further from state \emph{k}, the central plane wave part started to rise back, while the fronts also began to rise after a short decrease. Surprisingly, the pulse as a whole could not be totally blocked by the analyzer.

The observation clearly suggests that the DSR pulse does not acquire a single SOP. The SOP of the fronts differs from that of the central plane wave: the central plane wave acquires uniform elliptical polarization, whereas the fronts have spatially-varying polarization. Since the central plane wave part makes up the majority of the whole structure, when we experimentally aligned the axis of the analyzer at the angle of the minimal energy transmission, we approximately identified the minor axis of the central plane wave. Therefore, when the pulse then passed through the quarter-wave plate, only the central plane wave part was converted to linearly-polarized light. However, because the SOP of the fronts was different and spatially-varying, the fronts could not be converted to linearly-polarized light. As a result, when the pulse was finally analyzed by the analyzer, only the central plane wave could be fully blocked, whereas the fronts ended up as two separated oscillatory waves.

Afterwards, the behavior of the DSR pulse as pump power changes was investigated. Fixing the set-up at state $k$, we only changed the pump power. The distance between two fronts varied linearly with changes in the pump power. Figure.~\ref{fig:3b} shows typical results with pump powers of 300, 350, 400 $mW$. After the waveplate-analyzer, the extended central plane wave was always blocked, whereas two fronts show only small fluctuations. The plane wave and fronts maintain their unique polarization dynamics.

The experiment reveals that the vector soliton under DSR exhibits a non-uniform polarization distribution across the whole structure. The fronts and central plane wave feature different polarization dynamics. To confirm this, we performed numerical simulation on light circulation in the laser cavity. Pulse propagation through the fiber section is described by coupled Ginzburg-Landau equations having the form:

\begin{subequations}
\begin{eqnarray}
\frac{\partial u}{\partial z} &=& i \beta u - \sigma \frac{\partial u}{\partial t} - i \frac{k_2}{2} \frac{\partial^2 u}{\partial t^2} + i \frac{k_3}{6}\frac{\partial^3 u}{\partial t^3} + i \frac \gamma 3 v^2 u^* \nonumber \\
&&+ i \gamma \left(|u|^2+\frac 2 3 |v|^2\right)u + \frac g 2 u + \frac {g} {2 \Omega_g} \frac{\partial^2 u}{\partial t^2}  \\
\frac{\partial v}{\partial z} &=& -i \beta v + \sigma \frac{\partial v}{\partial t} - i \frac{k_2}{2} \frac{\partial^2 v}{\partial t^2} + i \frac{k_3}{6}\frac{\partial^3 v}{\partial t^3} + i \frac \gamma 3 u^2 v^* \nonumber \\
&& + i \gamma \left(|v|^2+\frac 2 3 |u|^2 \right)v + \frac g 2 v + \frac {g} {2 \Omega_g} \frac{\partial^2 v}{\partial t^2}
\end{eqnarray}
\end{subequations}
where $u$ and $v$ are the slowly-varying envelopes of the two linearly-polarized components along the slow and fast polarization axes of the optical fiber, respectively. Here $2\beta$ is the wave-number difference, $2\sigma$ is the inverse group velocity difference, $k_2, k_3, \gamma$ represent the second-order, third-order dispersion and Kerr nonlinearity of the fibers, $g$ is the saturable gain of the YDF and $\Omega_g$ is the gain bandwidth. A parabolic gain shape, with 40 $nm$ bandwidth, was assumed. The gain saturation was also considered for YDF
\begin{equation}
g = G_0 \exp \left[-\int \left(|u|^2 + |v|^2 \right)dt/E_{sat} \right]
\end{equation}
where $G_0$ is the small signal gain coefficient and $E_{sat}$ is the saturation energy. The saturable absorption of the NOLM was modeled by its transmission function \cite{Doran:88}
\begin{equation}
T = 1-2\alpha \left\{ 1+\cos \left[ \left( 1-2\alpha \right) \gamma L \left(|u|^2 + |v|^2 \right) \right] \right\}
\end{equation}
where $\alpha =0.2$ is the split ratio of the coupler. $L$ is the loop length of 7 $m$. To potentially match with the experimental conditions, we set the parameters as follows: $k_2 = 22~ps^2/km$, $k_3 = 0.1~ps^3/km$, $\gamma = 5.8~(W\cdot km)^{-1}$, $G_0 = 6.9~m^{-1}$, $E_{sat} = 5~nJ.$ Considering a weak birefringence cavity, the beat length of the fiber was set to 5 $m$. The numerical model was solved by the split-step Fourier method. The same stable solutions were reached when using different initial conditions.

A convenient way to analyse the polarization dynamics is to use the Stokes parameters, defined as
\begin{eqnarray}
S_0 = |u|^2 + |v|^2, &&~S_1 = |u|^2 - |v|^2,~S_2 = 2|u||v|cos \Delta \psi  \nonumber \\
&& S_3 = 2|u||v|sin \Delta \psi
\end{eqnarray}
where $\Delta \psi$ is the phase shift between two components. From the Stokes parameters, the orientation angle, $\phi$, of the polarization ellipse can be easily calculated as
\begin{equation}
\tan2\phi=\frac {S_2} {S_1}
\end{equation}

\begin{figure*}[!t]
\centering
\subfigure{
    \label{fig:4a}
    \includegraphics[width=2.0in]{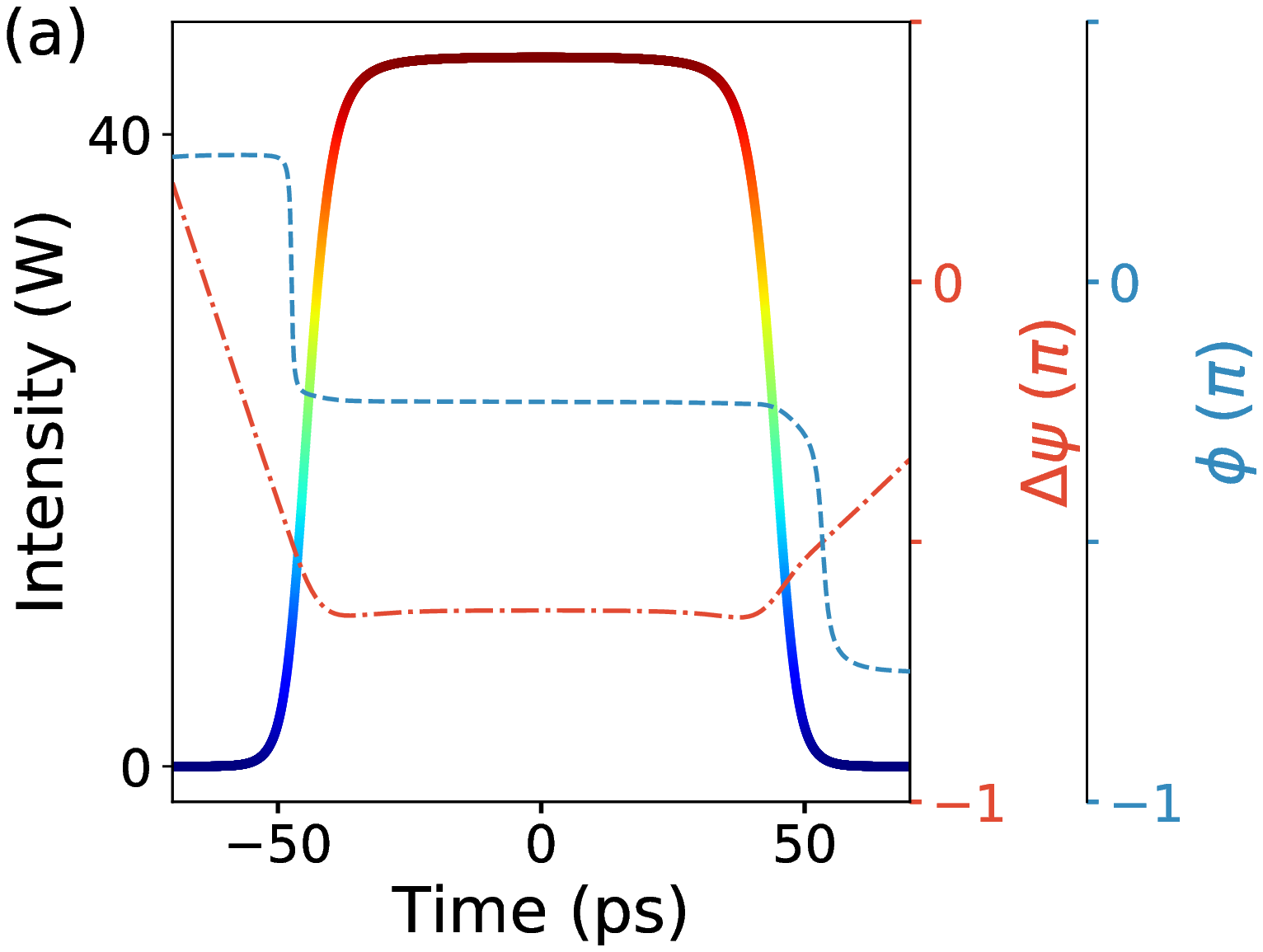}
}
\subfigure{
    \label{fig:4b}
    \includegraphics[width=2.0in]{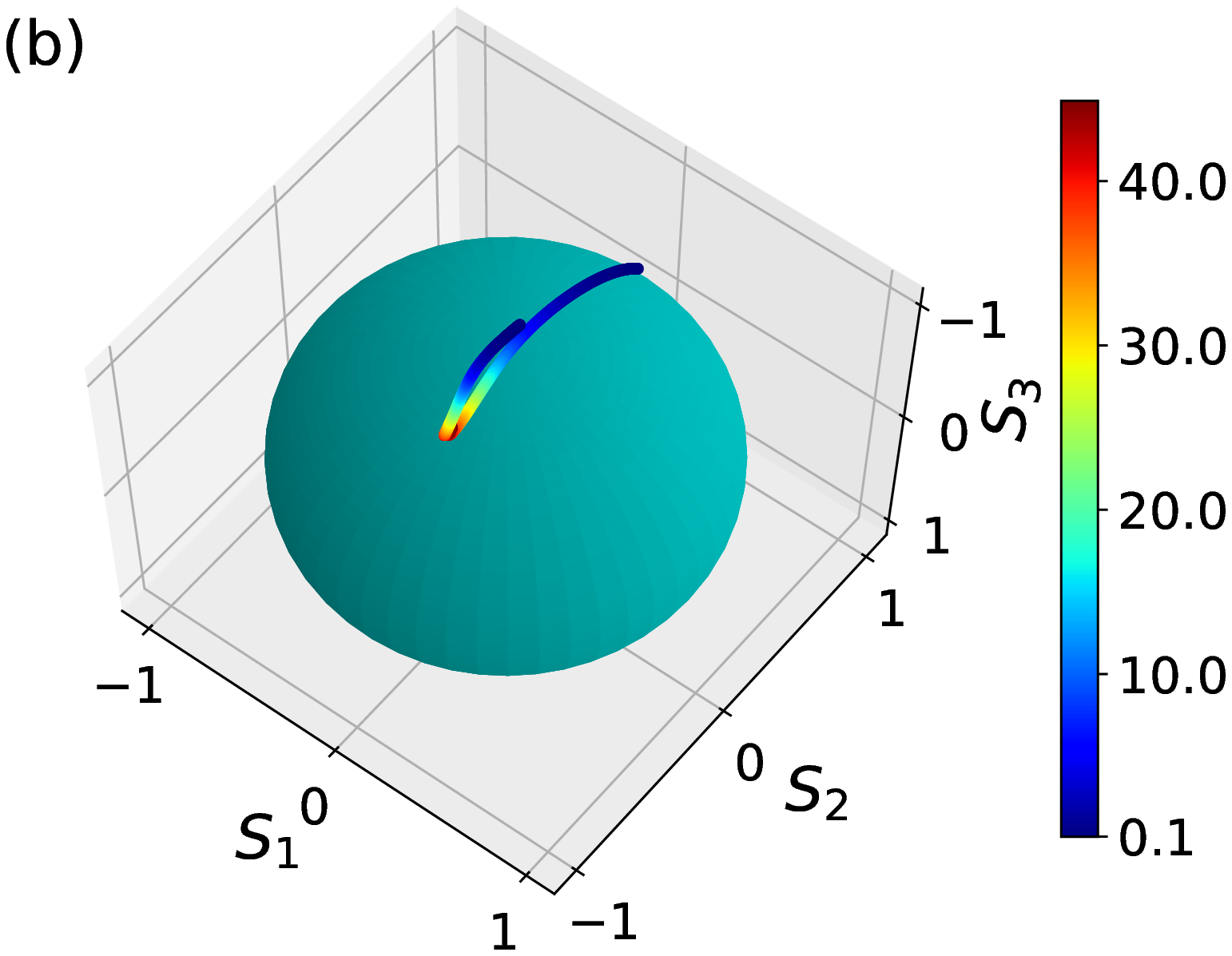}
}
\subfigure{
    \label{fig:4c}
    \includegraphics[width=1.9in]{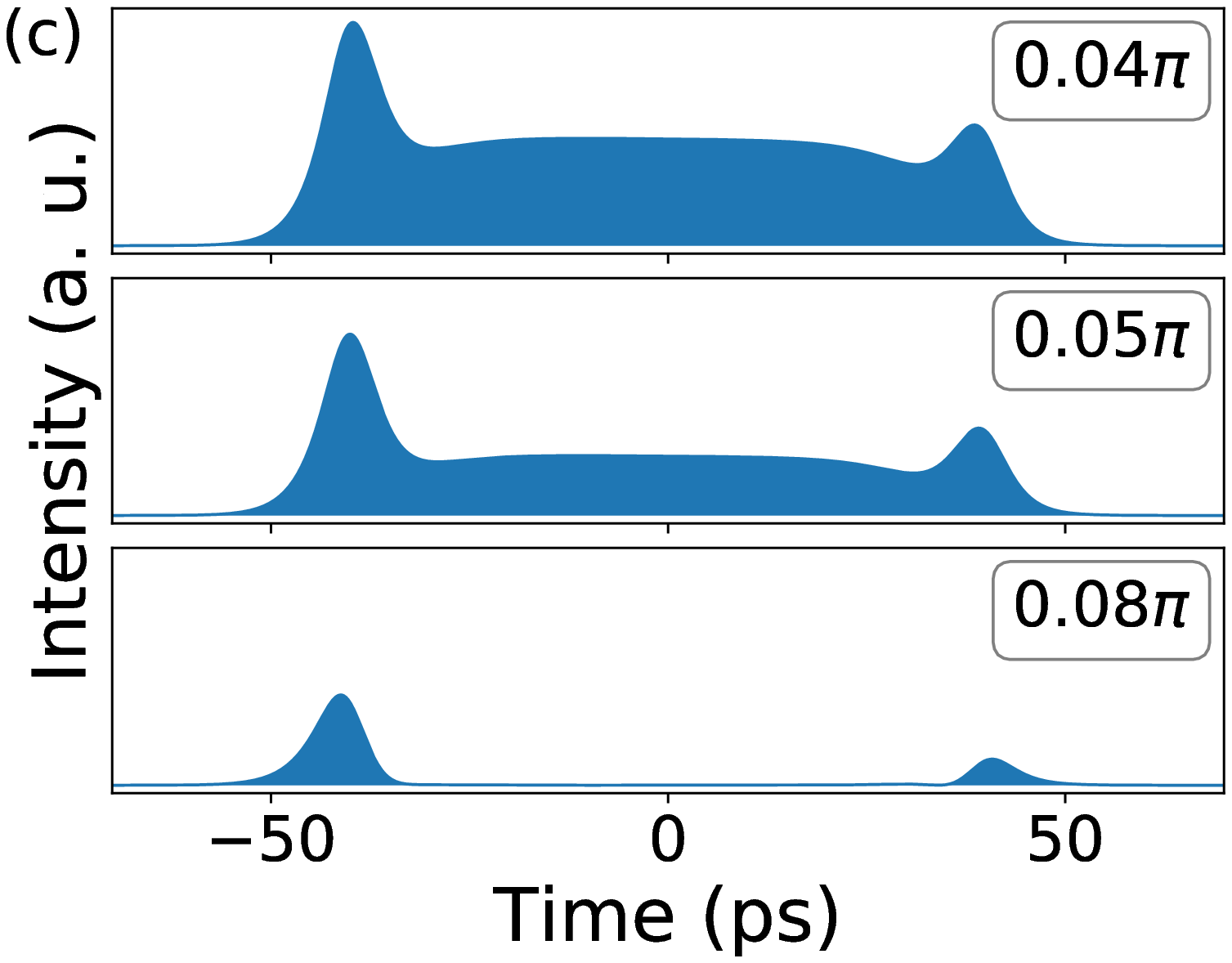}
}
\caption{Numerical results. (a) Colormap line: Calculated pulse, red dash dotted: the phase shift between its fast and slow components, blue dashed: orientation angle across the pulse. (b) Trajectory on the Poincaré sphere, of light field that meets $S_0(t) > 0.1$, color mapped by $S_0(t)$. (c) Simulated pulse profiles after the analyzer versus rotating transmission axis.}
\end{figure*}

In Fig.~\ref{fig:4a}, we plot the calculated resonant pulse and the phase shift $\Delta \psi(t)$ (red dash dotted) between its two components along the fast and slow axes of the fiber, as well as orientation angle $\phi(t)$ (blue dashed) across the pulse. The phase shift of two components is nearly the same across the central plane wave and starts to change when the fronts are approached. This also applies for the orientation angle of the polarization ellipse. The normalized Stokes parameters of the pulse are also plotted on the Poincar\'{e} sphere with color mapped by $S_0(t)$  in Fig.~\ref{fig:4b}. Note that only the part of the light wave that meets $S_0(t) > 0.1$ was plotted. One can see more clearly from the Poincar\'{e} sphere that the resonant pulse does not acquire a single SOP. The fronts show a long trajectory on the Poincar\'{e} sphere, while the central plane wave with the highest intensity converges to one point.

As the central plane wave forms the majority of the pulse, we verified that the largest energy ratio between the resolved components was found when the pulse was resolved along the major and minor axes of the polarization ellipse of the central plane wave. We then numerically perform the same polarization test, passing the pulse through the waveplate-analyzer combination.  The final fields after the analyzer, versus the rotation of the analyzer in the vicinity of the minimum energy transmission, are depicted in Fig. \ref{fig:4c}. This polarization non-conformity between fronts and the plane wave is maintained with varying gain settings. Comparing with the experiments in Fig.~\ref{fig:3}, the simulations show good agreement.

The experiment and simulation confirm that the central plane wave of the DSR pulse acquires a uniform SOP and that the fronts acquire spatially varying SOP. We believe that the fiber segment connecting the output coupler and the fiber collimator does not contribute to the distinct internal polarization distribution of the DSR pulse, for two reasons. Firstly, the length of this fiber segment is less then 1 $m$, and from the measured output energy and pulse width, the estimated pulse peak power is about $17.4~W$. Considering such a short fiber segment and the low peak power, this fiber segment would have a negligible nonlinear polarization rotating (NPR) effect on the pulse; this was then proved numerically. Secondly, we recall that the NPR effect is intensity-dependent. Intensities of the leading and trailing fronts of the original pulse are very close [see Fig.~\ref{fig:2}(a)]. However, the two fronts, after passing through the waveplate-analyzer, exhibit significant differences (see Fig.~\ref{fig:3}). Therefore, we believe that the DSR pulse acquires this internal polarization distribution within the mode-locked fiber laser. Meanwhile, the polarization states of the two fronts are asymmetric and are found to be dependent on the cavity birefringence. For example, in the case presented, the intensity of the leading front after the waveplate-analyzer was always higher than that of the trailing front. Experimentally, by rotating the intracavity PCs, different states could be observed where the trailing front was higher. This is also confirmed in the simulation, by simply changing the polarization parameter of the fiber.

In conclusion, we have observed vector DSR pulses in a mode-locked fiber laser. The DSR effect allows for a long soliton structure. With high-speed detection equipments, we are able to gain insight into the internal polarization dynamics across a long DSR pulse. Experimental results show that the DSR pulses do not acquire a single SOP. Although the central plane wave of the pulse acquires nearly a fixed polarization state, the fronts feature spatially-varying polarization states. Also, changes of gain condition do not affect the distinct polarization dynamics. Numerical simulations confirmed the experimental observations. Despite the polarization non-conformity between fronts and the plane wave, the DSR pulses are highly stable in the laser, suggesting that the composite balance between nonlinearity, dispersion and energy exchange forms a strong attractor in dissipative systems. The experimental and theoretical analysis gives important new insight into the internal dynamics of dissipative solitons.  We suggest that DSR pulses can form a good testbed for internal soliton dynamics investigations.

\begin{acknowledgments}
This work was supported in part by the National Natural Science Foundation of China under Grant 11674133, 11711530208, 61575089, and 61405079, in part by the Royal Society [grant number IE161214], in part by the Jiangsu Province Science Foundation under Grant BK20140231, in part by the Key Laboratory of Optoelectronic Devices and Systems of Ministry of Education and Guangdong Province. Daojing Li would like to thank Nail Akhmediev, Adrian Ankiewicz and Wonkeun Chang for fruitful discussions, and the China Scholarship Council for sponsorship.
\end{acknowledgments}


% \bibliography{ref.bib}
%

\end{document}